# Mass transport via in-plane nanopores in graphene oxide membranes

Tobias Foller[1], Lukas Madauß[2], Dali Ji[1], Xiaojun Ren[1], K. Kanishka H. De Silva[3], Tiziana Musso[1], Masamichi Yoshimura[3], Henning Lebius[4], Abdenacer Benyagoub[4], Priyank Kumar[5], Marika Schleberger[2,*], and Rakesh Joshi[1,*]

[1]School of Materials Science and Engineering, University of New South Wales, Sydney, NSW 2052, Australia.

[2]Faculty for Physics and CENIDE, University of Duisburg-Essen, 47057 Duisburg, Germany.

[3]Surface Science Laboratory, Toyota Technological Institute, Nagoya 468-8511, Japan.

[4]Normandie University, ENSICAEN, UNICAEN, CEA, CNRS, CIMAP, 14032 Caen, France

[5]School of Chemical Engineering, University of New South Wales, Sydney, NSW 2052, Australia

*Corresponding author. Email: marika.schleberger@uni-due.de, r.joshi@unsw.edu.au

Angstrom-confined solvents in 2-d laminates travel through interlayer spacings, gaps between adjacent sheets, and via in-plane pores. Among these, experimental access to investigate the mass transport through in-plane pores is lacking. Here, we create these nanopores in graphene oxide membranes via ion irradiation with precise control over functional groups, pore size and pore density. Low ion induced pore densities result in mild reduction and increased water permeation for the membranes. Higher pore densities lead to pronounced reduction and complete blockage of pure water however allows permeation of ethanol-water mixture due to weakening of hydrogen network. We confirm with simulations, that the attraction of the solvents towards the pores with functional groups and disruption of the angstrom-confined hydrogen network is crucial to allow in-plane pore transport.

Graphene oxide (GO) membranes are formed by stacking individual layers of functionalized graphene on top of each other[1]. The well-known tortuous pathway for mass transport within the interlayer network is formed (Fig. 1A)[1–7]. By creating in-plane nanopores, the water permeation through GO membranes can be enhanced by shortening the overall water pathway through the membrane[4,5] (Fig 1B and C). This is not only beneficial for water and organic solvent nano purification, but also for future battery technologies[8–10], electrochemical catalysts[11–14], supercapacitors[15,16] as well as biomedical technologies[17,18]. Despite being a crucial aspect for the mass transport, the transport mechanism through these in-plane nanopores in laminar 2-D materials is still not well investigated. Recent seminal studies focus mainly on the water or solvent flux through nanopores in single layer sheets or along pristine graphene channels [7,19–25]. However, as liquid is angstrom-confined within the laminar structures the transport mechanism inside 2-D laminar structures might be vastly different. Moreover, the recently discovered potential of GO membranes in organic nanofiltration calls for an investigation of the benefits of in-plane pores for organic solvent permeance in general[6,19,26].

In this study, we show that the mass transport via in-plane nanopores is influenced by the presence of functional groups as well as the surface tension of the permeating liquid. We show that the hydrogen network of water molecules inside the interlayer space is influenced by the presence of functional groups attached to the graphene. The disturbance of the water network allows the water to permeate faster. Moreover, by lowering the surface tension through the introduction of ethanol, the network is also disturbed and consequently the permeance is enhanced as well. This confirms the vital interplay between the liquid network strength and edge functionalities for the mass transport in 2-D laminar structures and supports theoretical predictions that edge functional groups govern the mass transport through nanopores in single layer sheets[21,27].

Ion irradiation is a well-established tool for tailoring materials on the nanoscale, in particular membranes and two-dimensional materials [28–30]. Here, ion irradiation is utilized to simultaneously create nanopores within the laminar GO structure and remove functional groups of the GO (Fig 1 A-C). That offers the chance to change the number of functional groups and create in-plane pores within a GO membrane and investigate the influence of both on the liquid mass transport through the GO laminar network. In order to avoid creating nanochannels perpendicular in respect to the membranes surface through the whole GO laminar network, the GO membranes were irradiated under a grazing incidence creating pores that are displaced from each other (Fig 1C). This further offers the chance to investigate the influence of the pore size on the mass transport as lowering the angle of incidence increases the pore size. Figure 1D and E exemplary show pores induced in monolayer GO by ion irradiation under 1° and 10°. For perpendicular incidence no detectable defects were found in the TEM analysis (see fig. S1). The ion irradiation under grazing incidence induces slit-like nano pores into GO monolayers, similar to those observed in suspended graphene and MoS$_2$ monolayers [29,32]. A statistical analysis of the pore sizes and shape (>50, >100 defects analysed for 1° and 10°) reveals that the average length is around 10 nm ± 4 nm for 10° and around 34 nm ± 14 nm for 1°. The average aspect ratio of the slit also increases from 1.8 ± 0.6 (for 10°) to 4.7 ± 2.5 (for 1°). The width of the slits are around 6 nm ± 2 nm (10°) and 8 nm ± 2 nm (1°). Hence, the length of the slit pores is adjustable while the width is less affected by the angle of incidence. In multilayer GO the length of the defects is further increased compared to monolayer as shown in fig. S2. The mechanism of pore creation under ion irradiation is discussed in more detail in SI-#1 and fig. S3.

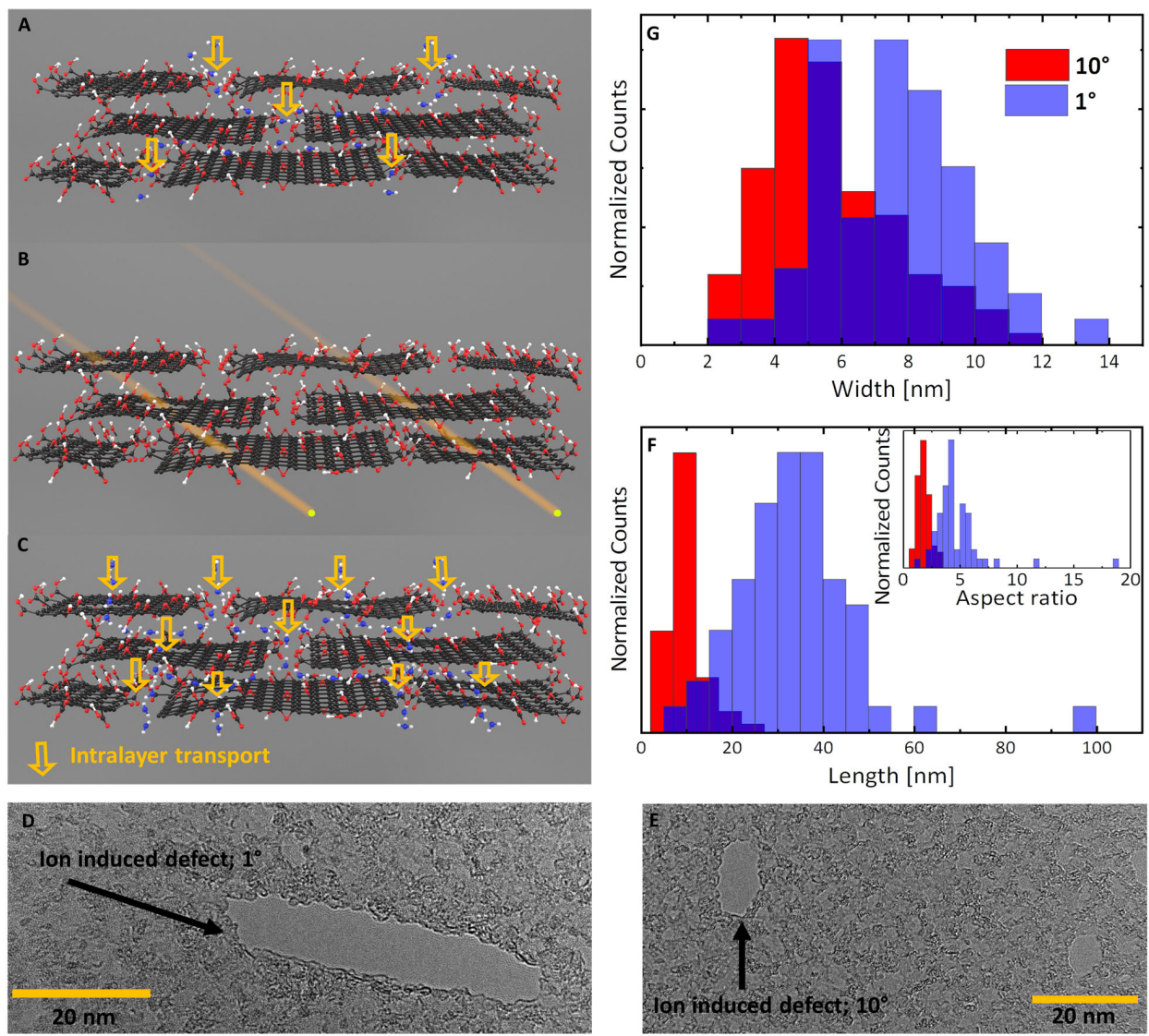

**Fig. 1 In-plane pore creation in GO via ion irradiation. (A)** Water transport in GO membranes. Intralayer transport (yellow arrows) is limited to sheet edges and intrinsic defects (not shown for clarity). **(B)** Grazing incidence in GO membranes create set-apart pores within laminar network. **(C)** Shortened tortuous pathway of mass transport in ion irradiated GO. The number of ions/cm$^2$ precisely controls the defect density. The utilized ions had a kinetic energy of ~90 MeV, so that the trajectory in the solid follows a straight line[31]. **(D-E)** TEM image of pores created under 1° and 10° grazing incidence in monolayer GO. **(F-G)** Statistical distribution of length and width of the pores created under grazing incidence. For the analysis >50 and >100 pores were measured. The inset shows the aspect ratio created for each individual defect.

In GO membranes, multiple GO layers are stacked on top of each other. As shown in the TEM analysis, increasing the number of layers changes the dimensions of the created pores. Thus, the pores created in GO membranes might be different from the defects created in monolayers. SEM images of the irradiated GO membranes indicate pore lengths between 10-100 nm (fig. S4A). XPS analysis of irradiated GO membranes show that with increasing fluence the C/O ratio is subsequently increased (fig. S4B, S5). The fluence necessary to increase the C/O ratio becomes less by decreasing the angle of incidence. This reduction is also confirmed by an increase in conductivity which follows the same trend as the C/O ratio in respect to fluence and angle of incidence (fig. S4C). Moreover, the reduction is further confirmed by Raman spectroscopy (fig. S4D). It may also be noted that the interlayer spacings of the GO membranes remain constant within less than 1 Å under ion irradiation (see fig. S6). The presence of the XRD peak also confirms that the layered structure remains intact after ion irradiation.

The chemical and structural analysis of the ion induced pores shows that the ion irradiation allows to control the density and size of in-plane pores as well as the reduction degree of the GO. With that it offers a unique platform to investigate the in-plane pore transport mechanism in GO membranes. Here the water permeance is tested for different pore density (fluence) and pore size (1° and 10° angle of incidence). Methylene Blue (MB) rejection was tested as a good indicator for whether the mass transport follows solely the in-plane pores or involves the tortuous pathway. As the length of the MB molecule (1.4 nm) is larger than the opening between the interlayer space but smaller than the ion induced pore width, it can pass through the ion induced in-plane pores, but not the tortuous interlayer network.

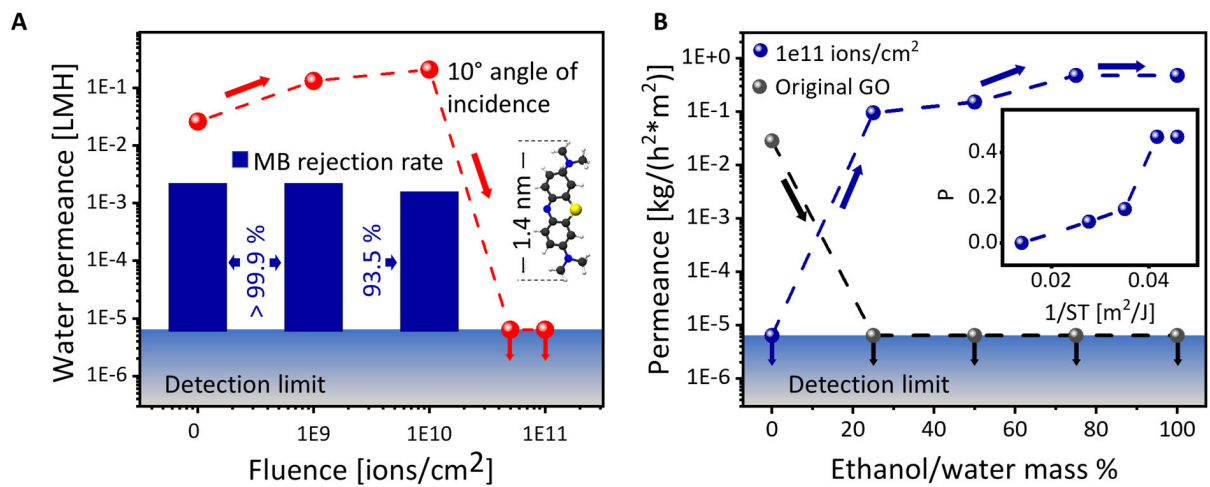

**Fig. 2 Mass transport through ion irradiated GO membranes. (A)** Water permeance in respect to ion fluence through GO membranes. The samples were irradiated under an angle of 10° (for 1° see fig. S10). The detection limit corresponds to the sensitivity of the scale (0.01 g) The histograms show the MB rejection rate. For samples with fluences > 7E10 ions/cm$^2$ no water permeated. Hence no MB rejection could be assessed. The displayed molecule is MB with a max length of 1.4 nm. **(B)** Permeance (P) of ethanol and water mixtures with varied ethanol concentrations through original GO and GO irradiated with 1E11 ions/cm$^2$ under 10°. The inset shows P of ethanol/water mixture in respect to the inverse surface tension (ST). Values for ST of water-ethanol mixture from [33].

Pressure-driven permeation experiments were conducted to assess the water transport and MB rejection (see fig S7, SI #3). As shown in fig. 2A, original GO membranes had a relatively low water permeance of 0.026 LMH, which is due to the high thickness of the samples. The cross-sectional SEM image reveals a thickness of around 1 μm (see fig. S8, SI #4). For 1E9 ions/cm$^2$ and 1E10 ions/cm$^2$ irradiated under 10° the water permeance gradually increases to 0.13 and 0.21 LMH corresponding to an increase by a factor of 5 and 8. Surprisingly, for higher pore densities, the water transport through the membranes was completely stopped (below the detection limit). Samples with fluences > 5e11 ions/cm$^2$ resulted in obvious microcracks in the membrane and are thus, not listed in the results. For original GO as well as 1E9 ions/cm$^2$ the level of MB in the permeate was below the detection limit (0.1 ppm). As the feed solution was 10 ppm MB, the rejection rate is > 99.9 %. Since the water permeance is increased and MB rejection retained it suggests that the mass transport follows in-plane pores as well as parts of the tortuous pathway. For 1E10 cm$^2$ the rejection rate slightly decreases to 93.5 %. This suggests that, for fluences >1E10 ions/cm$^2$, a fraction of the mass transport follows only in-plane pores instead of the tortuous pathway (see SI #5 and fig. S9 for detailed description of MB rejection rate determination).

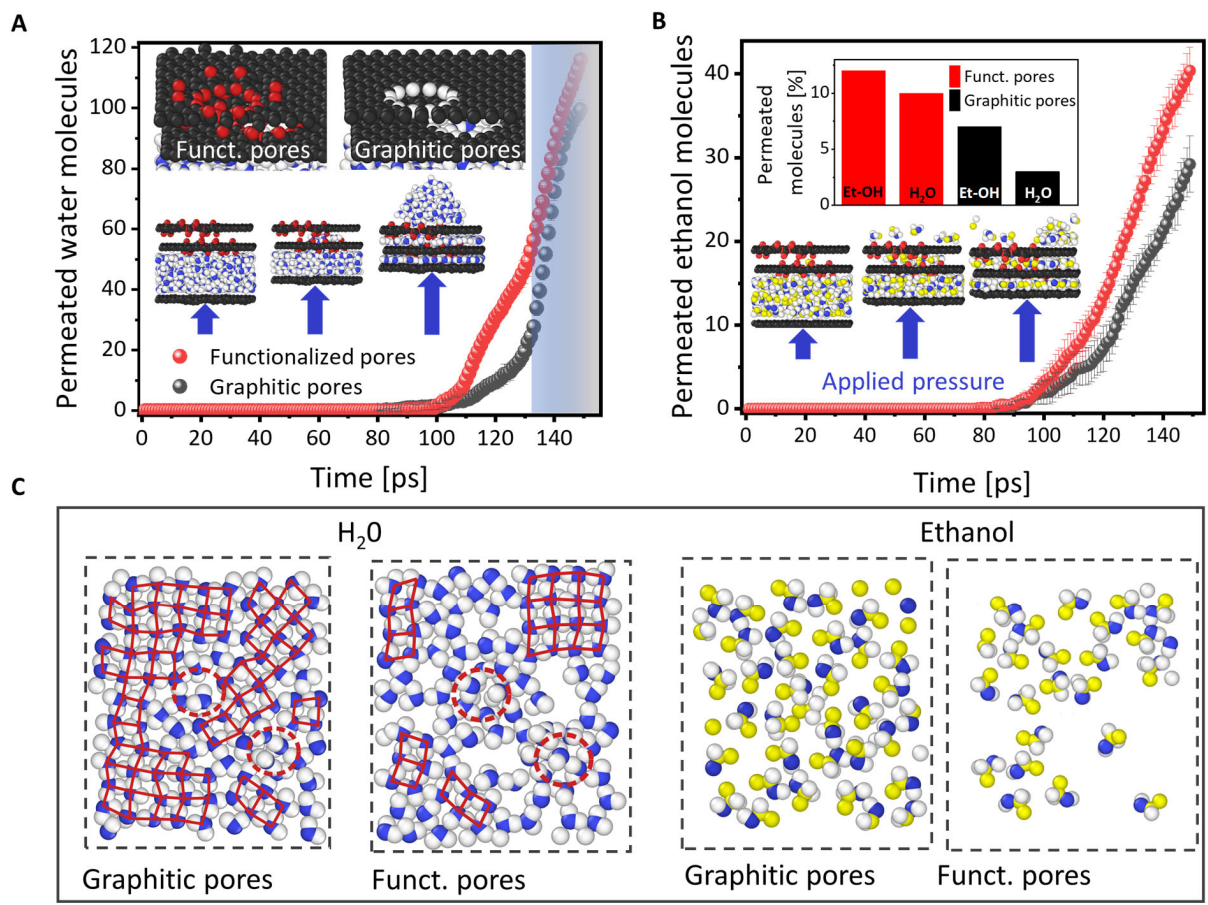

**Fig. 3. MD simulation of water/ethanol permeance through layered GO with graphitic and functionalized in-plane pores. (A-B)** Water/Ethanol permeance through respective holes. The rapid increase of water permeance after ~130 ps in graphitic pores is a simulation artefact as discussed in SI-#7 and fig S11. Inset in B shows percentage of permeated molecules after 120 ps. The pore diameter is around 10.2 Å. Additional simulations presented in SI - #7 **(C)** Snapshots of water and ethanol molecules in interlayer space at the last time interval of simulation. Square water network is marked with red lines, position of pores indicated with red dotted lines.

To further examine the mass transport through GO membranes with high pore densities and vanishing water permeance, the permeance of ethanol/water mixtures as well as pure ethanol was measured. Fig. 2B shows, that with increased ethanol content the permeance is increased and reaches a plateau for ethanol/water ratio >75 %. Further, the inset of fig. 2B shows that the permeance (P) increases with the inverse of the surface tension (ST). The concentration of ethanol and water was tested in the permeate side and no difference to the feed side was detected. As for original GO, ethanol/water mixtures > 25 % showed no detectable permeance through the membrane in over 72 h of constant applied pressure as previously reported for thick GO membranes[1,3].

In part the water permeance after ion irradiation follows an intuitive trend. With increasing pore densities and size the water transport is enhanced (see SI-#6 and Fig. S10 for 1 ° irradiation). However, for a certain number of pores, the water transport is decreased (1E9 ions/cm$^2$ under 1°, see fig. S10) or even stopped for samples irradiated under 10° and pore densities > 7E10 ions/cm$^2$ (see fig. 2A). With each created defect the GO is simultaneously reduced. The reduction of GO occurs in the surrounding areas of the defects presumably due to ion induced heating[34] (see also fig. S3). At higher pore densities, these reduced areas start to overlap with the neighbouring pores resulting in a higher removal of functional groups at the edge and surroundings of the pores. For example, as described in Fig. S2, the TEM images showing multilayer GO, illustrate the proximity of pores at densities around 5E10 ions/cm$^2$. This reduction may be responsible for changing the water transport across the membranes in total. For samples irradiated under 1° the reduction per created pore is even higher, thus the overlap of reduced areas and consequently the decrease of water transport occurs at lower defect densities (1E9 ions/cm$^2$).

To further explore the role of functional groups in the in-plane pore transport, molecular dynamics (MD) simulations were carried out. For that, the permeance of water and ethanol molecules through pores with and without functional groups in layered graphene sheets were simulated (see fig. 3 A-B and SI - #7). Pores without functional groups in their vicinity are named as graphitic pores. As shown in fig. 3A, the water permeance through graphitic pores is reduced compared to functionalized pores. The permeance of ethanol shows a similar trend as the functionalized pores exhibit a higher transport compared to the graphitic pores as well (fig. 3B). Comparing the percentage of permeated molecules at a certain time step (120 ps) reveals that ethanol permeates through the pores quicker than water (fig 3B, inset).

The dynamics of the water and ethanol molecules within the interlayer space allow to understand the differences in permeation of ethanol and water in functional and graphitic pores. Fig. 3C shows the configuration of the molecules in the interlayer space at the last time step of the simulation. More time steps are shown in fig. S12 and S13. It becomes apparent that water forms a network of squares through hydrogen interaction in the case of graphitic pores, as previously observed in 2-D angstrom-confined water [1,35]. That network is disturbed through the introduction of functional groups as they interact with the water molecules through hydrogen interaction. The disturbance of the network as well as the attraction of the water molecules towards the functional groups facilitates the water transport through the functionalized pores. In contrast, the ethanol molecules form no apparent network throughout the simulation in the interlayer space. Hence the formation of a network poses no competition to the permeation as compared to the case of water. This explains the enhanced permeance of ethanol through both functionalized and graphitic pores in the simulations. Additionally, the interaction between the functional group of ethanol and the functionalized pores enhances the transport compared to graphitic pores. These observations are nicely visualized in the full animation of the simulation in Video S1.

The simulation results agree well with the permeation experiments. Both show that functionalized in-plane pores in GO allow a fast intralayer transport. Moreover, the surface tension (i.e. the hydrogen network) plays a crucial role as well. This becomes apparent in the permeation experiment with a dependence of 1/ST of the permeance and is explained by the hindrance of the formation of a hydrogen network as observed in the MD simulations.

In conclusion, this study presents a novel way of tailoring GO membranes enabling us to gain important insights into the in-plane pore transport in graphene oxide membranes. Understanding the transport mechanism through pores in such laminates is essential for fine-tuning 2-D laminates for various applications [4–6,8–21,24,27]. Beyond that, our insights in the structural changes of interfacial water in the presence of functional groups may trigger further research in areas like the dielectric constant of water which may be dependent on the network formed under confinement [35].

## References

1. Nair RR, Wu HA, Jayaram PN, Grigorieva I v., Geim AK. Unimpeded permeation of water through helium-leak-tight graphene-based membranes. *Science*. 2012;335(6067):442-444. doi:10.1126/science.1211694

2. Huang L, Chen J, Gao T, et al. Reduced Graphene Oxide Membranes for Ultrafast Organic Solvent Nanofiltration. *Advanced Materials*. 2016;28(39):8669-8674. doi:10.1002/ADMA.201601606

3. Yang Q, Su Y, Chi C, et al. Ultrathin graphene-based membrane with precise molecular sieving and ultrafast solvent permeation. *Nature Materials 2017 16:12*. 2017;16(12):1198-1202. doi:10.1038/nmat5025

4. Ying Y, Sun L, Wang Q, Fan Z, Peng X. In-plane mesoporous graphene oxide nanosheet assembled membranes for molecular separation. *RSC Advances*. 2014;4(41):21425-21428. doi:10.1039/C4RA01495B

5. Saraswat V, Jacobberger RM, Ostrander JS, et al. Invariance of water permeance through size-differentiated graphene oxide laminates. *ACS Nano*. 2018;12(8):7855-7865. doi:10.1021/acsnano.8b02015

6. Nie L, Goh K, Wang Y, et al. Realizing small-flake graphene oxide membranes for ultrafast size-dependent organic solvent nanofiltration. *Science Advances*. 2020;6(17). doi:10.1126/SCIADV.AAZ9184

7. Abraham J, Vasu KS, Williams CD, et al. Tunable sieving of ions using graphene oxide membranes. *Nature Nanotechnology*. 2017;12(6):546-550. doi:10.1038/nnano.2017.21

8. Zhang L, Yue J, Wei T, et al. Densely pillared holey-graphene block with high-level nitrogen doping enabling ultra-high volumetric capacity for lithium ion storage. *Carbon*. 2019;142:327-336. doi:10.1016/J.CARBON.2018.10.070

9. Jin Y, Hu C, Dai Q, et al. High-Performance Li-CO2 Batteries Based on Metal-Free Carbon Quantum Dot/Holey Graphene Composite Catalysts. *Advanced Functional Materials*. 2018;28(47):1804630. doi:10.1002/ADFM.201804630

10. Zhao X, Hayner CM, Kung MC, Kung HH. Flexible Holey Graphene Paper Electrodes with Enhanced Rate Capability for Energy Storage Applications. *ACS Nano*. 2011;5(11):8739-8749. doi:10.1021/NN202710S

11. Yan H, Xie Y, Jiao Y, et al. Holey Reduced Graphene Oxide Coupled with an Mo2N–Mo2C Heterojunction for Efficient Hydrogen Evolution. *Advanced Materials*. 2018;30(2):1704156. doi:10.1002/ADMA.201704156

12. Kong W, Yao K, Duan X, Liu Z, Hu J. Holey Co, N-codoped graphene aerogel with in-plane pores and multiple active sites for efficient oxygen reduction. *Electrochimica Acta*. 2018;269:544-552. doi:10.1016/J.ELECTACTA.2018.02.148

13. Deng D, Novoselov KS, Fu Q, Zheng N, Tian Z, Bao X. Catalysis with two-dimensional materials and their heterostructures. *Nature Nanotechnology*. 2016;11(3):218-230. doi:10.1038/nnano.2015.340

14. Foller T, Daiyan R, Jin X, et al. Enhanced graphitic domains of unreduced graphene oxide and the interplay of hydration behaviour and catalytic activity. *Materials Today*. Published online September 9, 2021. doi:10.1016/J.MATTOD.2021.08.003

15. Han X, Funk MR, Shen F, et al. Scalable Holey Graphene Synthesis and Dense Electrode Fabrication toward High-Performance Ultracapacitors. *ACS Nano*. 2014;8(8):8255-8265. doi:10.1021/NN502635Y

16. Xu Y, Chen CY, Zhao Z, et al. Solution Processable Holey Graphene Oxide and Its Derived Macrostructures for High-Performance Supercapacitors. *Nano Letters*. 2015;15(7):4605-4610. doi:10.1021/ACS.NANOLETT.5B01212

17. Salah A, Al-Ansi N, Adlat S, et al. Sensitive nonenzymatic detection of glucose at PtPd/porous holey nitrogen-doped graphene. *Journal of Alloys and Compounds*. 2019;792:50-58. doi:10.1016/J.JALLCOM.2019.04.021
18. Lokhande AC, Qattan IA, Lokhande CD, Patole SP. Holey graphene: an emerging versatile material. *Journal of Materials Chemistry A*. 2020;8(3):918-977. doi:10.1039/C9TA10667G
19. Cheng C, Iyengar SA, Karnik R. Molecular size-dependent subcontinuum solvent permeation and ultrafast nanofiltration across nanoporous graphene membranes. *Nature Nanotechnology 2021 16:9*. 2021;16(9):989-995. doi:10.1038/s41565-021-00933-0
20. Surwade SP, Smirnov SN, Vlassiouk I v., et al. Water desalination using nanoporous single-layer graphene. *Nature Nanotechnology*. 2015;10(5):459-464. doi:10.1038/nnano.2015.37
21. Cohen-Tanugi D, Grossman JC. Water Desalination across Nanoporous Graphene. *Nano Letters*. 2012;12(7):3602-3608. doi:10.1021/NL3012853
22. Gopinadhan K, Hu S, Esfandiar A, et al. Complete steric exclusion of ions and proton transport through confined monolayer water. *Science (New York, NY)*. 2019;363(6423):145-148. doi:10.1126/science.aau6771
23. Esfandiar A, Radha B, Wang FC, et al. Size effect in ion transport through angstrom-scale slits. *Science*. 2017;358(6362):511-513. doi:10.1126/SCIENCE.AAN5275/SUPPL_FILE/AAN5275_ESFANDIAR_SM.PDF
24. Radha B, Esfandiar A, Wang FC, et al. Molecular transport through capillaries made with atomic-scale precision. *Nature*. 2016;538(7624):222-225. doi:10.1038/nature19363
25. Holt JK, Park HG, Wang Y, et al. Fast mass transport through sub-2-nanometer carbon nanotubes. *Science*. 2006;312(5776):1034-1037. doi:10.1126/SCIENCE.1126298/SUPPL_FILE/HOLT.SOM.PDF
26. Zheng S, Tu Q, Wang M, Urban JJ, Mi B. Correlating Interlayer Spacing and Separation Capability of Graphene Oxide Membranes in Organic Solvents. *ACS Nano*. 2020;14(5):6013-6023. doi:10.1021/acsnano.0c01550
27. Wang L, Boutilier MSH, Kidambi PR, Jang D, Hadjiconstantinou NG, Karnik R. Fundamental transport mechanisms, fabrication and potential applications of nanoporous atomically thin membranes. *Nature Nanotechnology 2017 12:6*. 2017;12(6):509-522. doi:10.1038/nnano.2017.72
28. Madauß L, Schumacher J, Ghosh M, et al. Fabrication of nanoporous graphene/polymer composite membranes. *Nanoscale*. 2017;9(29):10487-10493. doi:10.1039/C7NR02755A
29. Ochedowski O, Lehtinen O, Kaiser U, et al. Nanostructuring graphene by dense electronic excitation. *Nanotechnology*. 2015;26(46):465302. doi:10.1088/0957-4484/26/46/465302
30. Pérez-Mitta G, Toimil-Molares E, Trautmann C, et al. Molecular Design of Solid-State Nanopores: Fundamental Concepts and Applications. *Advanced Materials*. 2019;31(37):1901483. doi:10.1002/ADMA.201901483
31. Akcöltekin E, Peters T, Meyer R, et al. Creation of multiple nanodots by single ions. *Nature Nanotechnology 2007 2:5*. 2007;2(5):290-294. doi:10.1038/nnano.2007.109

32. Madauß L, Zegkinoglou I, Vázquez Muiños H, et al. Highly active single-layer MoS2 catalysts synthesized by swift heavy ion irradiation. *Nanoscale*. 2018;10(48):22908-22916. doi:10.1039/C8NR04696D

33. Vázquez G, Alvarez E, Navaza JM. Surface Tension of Alcohol + Water from 20 to 50 °C. *J Chem Eng Data*. 1995;40:611-614. Accessed November 8, 2021. https://pubs.acs.org/sharingguidelines

34. Tyagi C, Khan SA, Sulania I, Meena R, Avasthi DK, Tripathi A. Evidence of Ion-Beam-Induced Annealing in Graphene Oxide Films Using in Situ X-Ray Diffraction and Spectroscopy Techniques. *Journal of Physical Chemistry C*. 2018;122(17):9632-9640. doi:10.1021/acs.jpcc.7b10699

35. Fumagalli L, Esfandiar A, Fabregas R, et al. Anomalously low dielectric constant of confined water. *Science*. 2018;360(6395):1339-1342. doi:10.1126/SCIENCE.AAT4191/SUPPL_FILE/AAT4191_FUMAGALLI_SM.PDF

36. You Y, Jin XH, Wen XY, et al. Application of graphene oxide membranes for removal of natural organic matter from water. *Carbon*. 2018;129:415-419. doi:10.1016/j.carbon.2017.12.032

37. Ochedowski O, Osmani O, Schade M, et al. Graphitic nanostripes in silicon carbide surfaces created by swift heavy ion irradiation. *Nature Communications 2014 5:1*. 2014;5(1):1-8. doi:10.1038/ncomms4913

38. Bradder P, Ling SK, Wang S, Liu S. Dye Adsorption on Layered Graphite Oxide. *Journal of Chemical and Engineering Data*. 2010;56(1):138-141. doi:10.1021/JE101049G

39. Plimpton S. Fast Parallel Algorithms for Short-Range Molecular Dynamics. *Journal of Computational Physics*. 1995;117(1):1-19. doi:10.1006/JCPH.1995.1039

40. Jorgensen WL, Maxwell DS, Tirado-Rives J. Development and testing of the OPLS all-atom force field on conformational energetics and properties of organic liquids. *Journal of the American Chemical Society*. 1996;118(45):11225-11236. doi:10.1021/JA9621760

41. Jewett AI, Stelter D, Lambert J, et al. Moltemplate: A Tool for Coarse-Grained Modeling of Complex Biological Matter and Soft Condensed Matter Physics. *Journal of Molecular Biology*. 2021;433(11):166841. doi:10.1016/J.JMB.2021.166841

42. Mark P, Nilsson L. Structure and Dynamics of the TIP3P, SPC, and SPC/E Water Models at 298 K. Published online 2001. doi:10.1021/jp003020w

43. Martinez L, Andrade R, Birgin EG, Martínez JM. PACKMOL: A package for building initial configurations for molecular dynamics simulations. *Journal of Computational Chemistry*. 2009;30(13):2157-2164. doi:10.1002/JCC.21224

44. Stukowski A. OVITO---the Open Visualization Tool. Published online 2010. doi:10.1088/0965-0393/18/1/015012

45. Ferrari AC, Robertson J. Interpretation of Raman spectra of disordered and amorphous carbon. *Physical Review B*. 2000;61(20):14095. doi:10.1103/PhysRevB.61.14095

**Acknowledgements**

T.F. acknowledges the UNSW Scientia Ph.D. Scholarship. R.J acknowledges the fellowship from Alexander-von-Humboldt-Foundation. T.F, L.M, R.J and M.S would like to acknowledge support by the RADIATE project under the Grant Agreement 824096 from the EU Research and Innovation programme HORIZON 2020. The authors also thank the facility of GANIL, France where the ion irradiation in this study was performed. The authors acknowledge the facilities and the scientific and technical assistance of Microscopy Australia at the Electron Microscope Unit (EMU) and Solid State & Elemental Analysis Unit (SSEAU) within the Mark Wainwright Analytical Centre (MWAC) at UNSW Sydney and at The University of Sydney. M.S. acknowledges financial support from the German Research Foundation (DFG) by funding SCHL 384/16-1 (project number 279028710).**Competing interests:** Authors declare that they have no competing interests.

**Data and materials availability:** All data are available in the main text or the supplementary materials except for all TEM images used for characterizing defect size. Request for this data should be addressed to RJ.

# Supplementary Information

# Mass transport via in-plane nanopores in graphene oxide membranes

Tobias Foller[1], Lukas Madauß[2], Dali Ji[1], Xiaojun Ren[1], K. Kanishka H. De Silva[3], Tiziana Musso[1], Masamichi Yoshimura[3], Henning Lebius[4], Abdenacer Benyagoub[4], Priyank Kumar[5], Marika Schleberger[2,*], and Rakesh Joshi[1,*]

[1]School of Materials Science and Engineering, University of New South Wales, Sydney, NSW 2052, Australia.

[2]Faculty for physics and CENIDE, University of Duisburg-Essen, 47057 Duisburg, Germany.

[3]Surface Science Laboratory, Toyota Technological Institute, Nagoya 468-8511, Japan.

[4]Normandie University, ENSICAEN, UNICAEN, CEA, CNRS, CIMAP, 14032 Caen, France

[5]School of Chemical Engineering, University of New South Wales, Sydney, NSW 2052, Australia

*Corresponding author. Email: marika.schleberger@uni-due.de, r.joshi@unsw.edu.au

## Materials and Methods

**GO Membrane and TEM grid preparation**

GO membranes were prepared as described in our previous studies [36]. A stock solution of GO (0.1 mg/ml) synthesized by Hummer's method. This solution was used to fabricate GO membranes via vacuum filtration through a PVDF membrane (Area: ~3.0 cm$^2$, 0.2 µm pore size) by applying a pressure of 60 kPa. Throughout the study the same amount of GO was used for each membrane to ensure a constant thickness. For the preparation of TEM grids, a highly diluted and sonicated GO solution was drop casted onto lacy carbon TEM grids. Before further treatment or analysis, the samples were stored in a desiccator.

**Ion irradiation**

The ion beam irradiation was done at IRRSUD of the GrandAccelerateur National d'Ions Lourds (GANIL) in Caen, France. The samples were irradiated homogenously across the surface with $_{136}$Xe$^{23+}$ ions with an energy of 0.71 MeV/u . For grazing incidence, the same methods as described in previous studies was used to calibrate the angle of incidence [31,37]. The angle was further validated by irradiation of an SrTiO3 crystal under several angle of incidences. With atomic force microscopy (AFM) the defect density was calculated assuming every ion causes a defect. Using the geometrical relation between angle of incidence and fluence the calibration of the angle could be verified within +-0.3°. The GO membranes were pressed against the sample holder with a custom designed set-up that allows to make the surface of the membrane as flat as possible. TEM grids were mounted on the sample holder for irradiation. The ion flux was kept below 1E9 ions/cm$^2$ * s$^{-1}$ to avoid heating of the sample.

**Water and ethanol flux**

Filtration measurements were conducted in a Sterlitech stirred cell (HP4750 Stirred Cell). The pressure vessel was filled with DI water or ethanol/water mixture as given in the results section above. During the filtration experiments 2 bar pressure was applied. As described in fig. S7 the permeate is collected into a vessel. The weight of the vessel is recorded every 10 s to measure the flux. The measurement of the flux was performed for at least 24 h to allow a stabilizing of the flux. Typically, the flux is initially much higher in the first 10 h. The recorded values shown in this study represent the stabilized flux after at least 15 h of filtration. If the recorded flux was reported as below detection limit, the samples were tested for at least 72 h and no water permeated during this time.

**Rejection test**

MB rejection was tested in the same set-up as the water and ethanol flux. To ensure that the MB rejection rates represent the state of the membrane as tested in the flux measurements, the MB rejection test were undertaken immediately after the flux measurements without changing the membrane sealing. As feed, 0.1 M MB water solution was filled into the Stirred cell. After at least 24 h of continuous filtration the feed side was analysed with UV-Vis spectrometer (PerkinElmer Lambda 365). By calibrating the UV-vis spectra of the adsorption with solutions of know concentration, the concentration of the permeate samples can be measured from their adsorption spectra. By using the following equation, the rejection rate R can be calculated as follows:

$$R = 1 - C_P/C_F$$

$C_F$ is the concentration of the feed and $C_p$ the concentration of the permeate side. MB might also get adsorbed within the laminar network of GO[38]. For that reason we let the filtration run for at least 24 h to make sure the potential adsorption sites get saturated. According the data in ref. [38] the adsorption capacity of MB on GO is around 200 mg/g. The GO membranes in this contain ~1.2 mg GO. Hence, the adsorption capacity of the GO membranes is around 0.24 mg MB. The MB solution used for filtration here has a concentration of 31.985 mg/ml. After 24 h we usually got several ml of permeate. Therefore the error in rejection rate due to the adsorption of MB should be neglectable even if the adsorption capacity of the GO used in this study might be different to the one in Ref. [38].

Ethanol content in the permeate side was analysed by testing the total organic carbon (TOC) content (Multi NC from Analytic Jena, Germany). A comparison with the TOC of the feed side allows to estimate a change in concentration during the filtration.

**MD Simulations**

All molecular dynamics (MD) simulations were conducted using the LAMMPS package [39]. The units in our simulations were set to "real". Prior to conducting MD simulations, we performed energy minimization using the steepest descent method with an energy tolerance of 1E-7 and a force tolerance of 1E-9. Following this, GO was fixed during our simulations and other molecules were allowed to move. Specifically, in the next step, the velocities for all the movable atoms were assigned a Maxwell-Boltzmann distribution at 300 K. A 50 ps NVT ensemble was employed to get the equilibrium systems. And then, another NVT ensemble was used to simulate molecular motion through GO pores. Here, the classical equations of motion were integrated with a time step of 1 fs and the simulation ran for a duration of 150 ps. During this phase, the lower graphene wall was allowed to move and pushed towards to upper GO wall with the pore with a velocity of 0.0001 Å/fs in order to facilitate the passage of molecules through the pore and study the resulting effects. We used periodic boundary conditions in all directions. A large vacuum space was taken into account in the z-direction, normal to the GO sheets, in order to prevent molecules that pass through the pore from interacting with the neighboring image molecules in the z-direction.[40,41]

The Particle-Particle Particle-Mesh method was used to evaluate the long-range electrostatic interactions. GO was described using the ReaxFF potentials, while the OPLSAA force fields and the SPC model [42] were used for ethanol and water, respectively. The interaction between GO and ethanol/water was described using the standard 12-6 Lennard-Jones (LJ) potential. The inter-atomic LJ parameters between species i and j were calculated using Lorentz-Berthelot mixing rules. The GO model parameters are given below:

**Table S1: Parameters for the Lennard-Jones (LJ) potential used for GO when modeling the GO-molecule (ethanol/water) interaction.**

| Atom | type | σ (Å) | ε (kJ/mol) |
|---|---|---|---|
| C | graphene | 3.43 | 0.44 |
|   | epoxy |  |  |
| O | epoxy | 3.12 | 0.25 |

Structures for MD simulations

Two graphene sheets with 20 Å spacing between them were created in a unit cell measuring 29.54 Å x 25.58 Å in the x-y directions. Holes (small and large relatively) were made in the upper graphene sheet and the hole was oxygenated with epoxy groups. For the oxidized graphene case, we added additional epoxy groups surrounding the hole. We filled the 20 Å space with 110 ethanol molecules or 310 water molecules using Packmol [43]. In order to model bilayer GO sheets, we added an extra GO layer with a hole (similar to the one described previously), but shifted 6 Å each in x and y directions. The interlayer spacing was also kept at 6 Å. Ovito [44] was used for trajectory visualization and analysis purposes as well as to visualize video S1.

**Characterization**

TEM measurements were conducted in a JOEL F200 at 80 keV and images were taken with a Gatan OneView s camera. The low acceleration voltage was chosen to avoid knock-on damage to the GO. XRD patterns were obtained from an Empyrean Thin-Film XRD. XPS measurements were collected out on an ULVAC-PHI 5000 Versa probe II, with an Al-Kα monochromatic X-ray source (energy = 1486.68 eV). As a binding energy reference, C1s = 284.8 eV for adventitious hydrocarbon was used. Raman spectra were recorded on a Renishaw in-via2 spectrometer (532 nm). The instrument was calibrated with a Silica surface. For each data point a Raman map of several 10 µm$^2$ and > 100 spectra was recorded, analysed and the given height ratios of the peak are averaged over those spectra. The sheet resistance was measured with a two-probe geometry using spring electrodes and a Keithley 2400 source meter. At least 4 different spots on each sample were measured and averaged. SEM images were generated using a FEI Nova NanoSEM 230 FE-SEM.

**Additional TEM images**

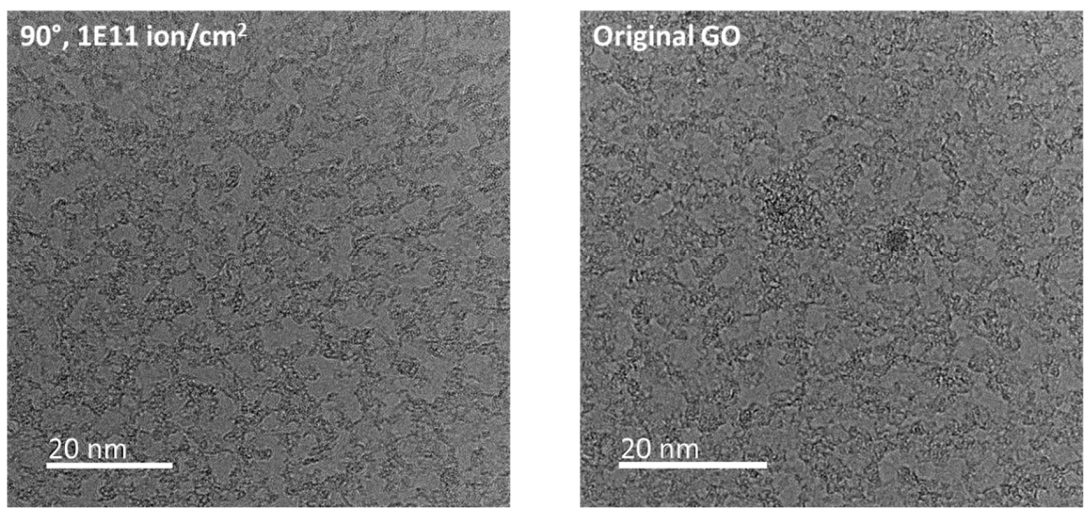

**Fig. S1 TEM image of original GO and GO irradiated under 90°.** For samples irradiated under 90° no apparent defect could be found. As a control sample original GO without irradiation is also shown

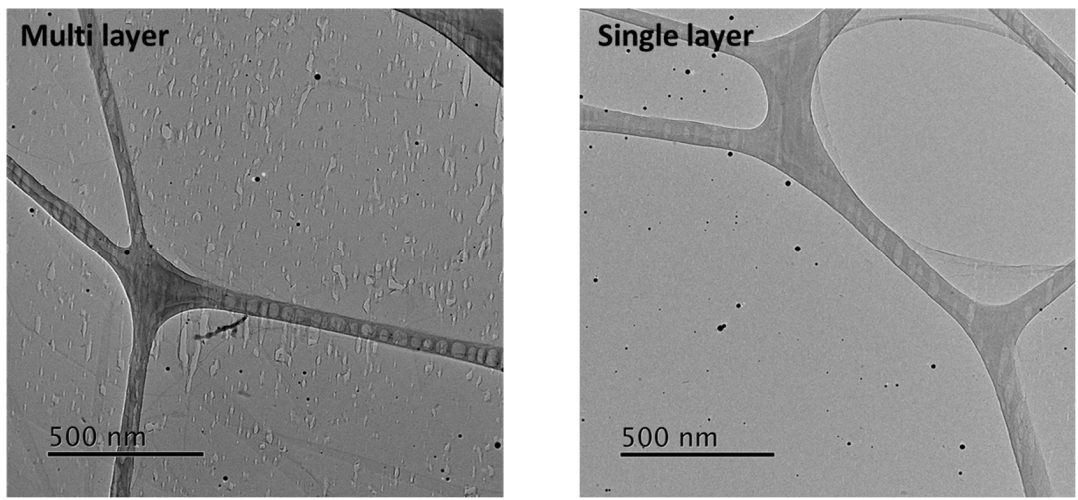

**Fig. S2 TEM images of defects in multi- and single layer.** Images show a comparison between defect size in multilayer and single layer GO. The images were taken on the same TEM grid to ensure that the irradiation conditions were the same. The defect size increase for multi-layer compared to single layer. The TEM grid was irradiated with 5E10 ions/cm$^2$ under 10° angle of incidence. Black dots are due to containments during the preparation process.

**Supplementary Text**

**#1 - Defect creation and reduction upon grazing incidence of graphene oxide**

As described in the main text, the irradiation of GO with high energy ion leads to defects and reduction. Further it was shown that the defect size, especially length and reduction degree per ion is dependent on the angle of incidence. Here we can utilize the core-halo model for the ion track with the numbers for GO derived from Tyagi et al to explain the general trend of angle dependent reduction and defect creation[34]. As an ion hits the surface of GO the temperature around the impact point increases sharply. Away from the centre of the impact, the temperature decreases radially. This allows to divide the area around the impact in two areas. One area closest to the impact spot, the core, reaches temperatures above the melting point of GO. The second area further away from the impact, the halo, reaches temperatures high enough to remove functional groups from GO[34]. It may also be noted that Tyagi et al investigated ion irradiation of GO in a stopping power range similar to this study. They used 1.51 keV/A electronic stopping power whereas this study utilizes 1.25 keV/A as computed for a C, O (ratio 2:1 and density of ~ 2 g/cm$^3$) target with the computer code SRIM.

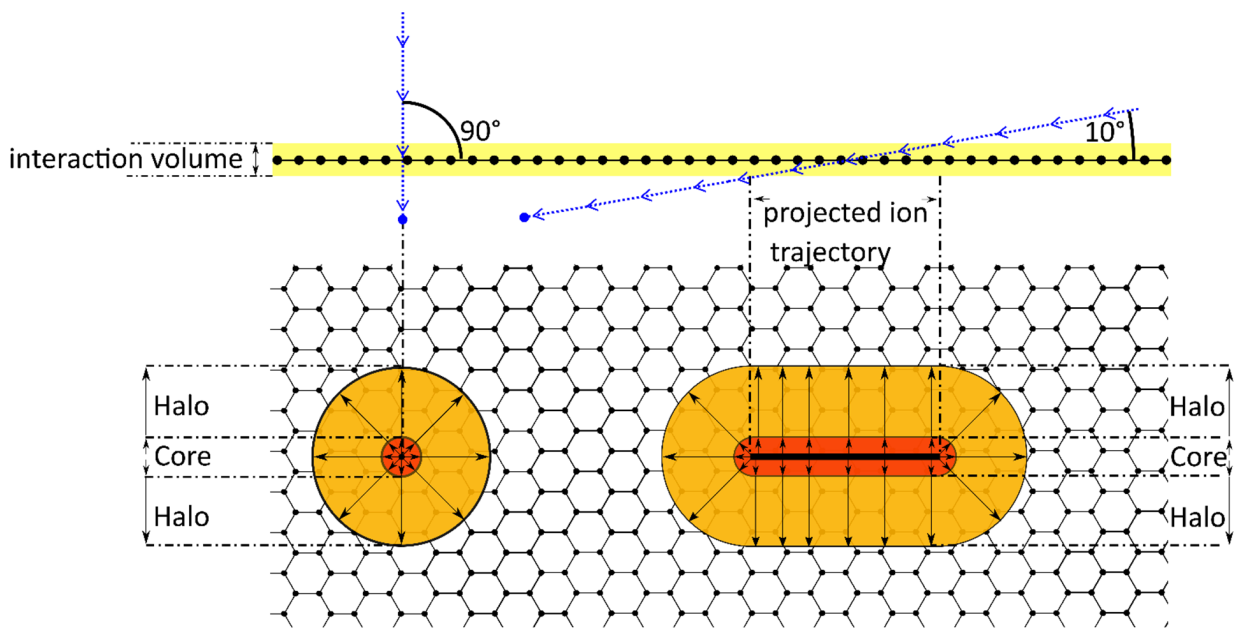

**Fig. S3 Geometric considerations for defect creation and reduction of GO upon ion irradiation under grazing incidence. Top:** Sideview of a GO sheet. Black dots and lines represent the graphene lattice, functional groups are not shown for better overview. Yellow area represents the volume of the electronic system around the GO lattice which interacts with ions that hit the surface. The blue dotted line resembles the trajectory of an ion (blue dot) hitting the GO surface under 90° (left) and 10° (right) in respect to the surface. **Bottom:** Top view of a GO sheet after ion irradiation. Functional groups are not shown for a better overview. Core and Halo areas represent temperatures above the melting point of GO and above the temperature needed to remove functional groups. By projecting the trajectory of the ion through the interaction volume, the core and halo model gives a good explanation for the formation of slit-like pores with angle-dependent length and reduction degree.

Fig. S3 shows simple geometrical considerations, that may explain the formation of slit-like pores, the angle dependence of the slit length and the increased reduction with smaller angles of incidences. These considerations are similar to models from other studies under grazing incidence [31,32]. When a high energetic ion (blue dot) hits the GO surface (black stick and ball), energy is transferred to the electronic system via electronic stopping. For simplicity, the electronic density of GO is considered constant around the GO lattice, this gives an interaction volume in which the ion interacts with the electronic system while passing through the GO plane (yellow area). When the ion hits the GO plane perpendicular, the heat is induced in a single impact spot from which the energy spreads in an isotropic manner from the point of impact. Thus, the resulting defect and reduced area in GO is circular as described by Tyagi et al[34].

If the ion hits the GO under a smaller angle of incidence, for example 10° in this study, the projected trajectory becomes a slit as shown in fig. S3. As the heat spreads in an isotopic manner from each point of the slit, the resulting core has an elongated slit shape. According to this model, the width of the core under 10° irradiation should be similar to the diameter of the core in perpendicular incidence and independent of further decrease of the angle. The predicted core diameter by Tyagi et al. is 8 nm. For 10° and 1° the measured width of the pores is 4-5 and 5-8 nm respectively and are thus in good agreement considering that the stopping power used in Tyagi's study is slightly higher. The slight increase in width between 10° and 1° might be explained by the fact that the overall cross-section between ion and interaction volume increases for smaller angles, allowing a longer interaction and thus higher energy transfer between the target's electronic system and the ion. Additionally, deviations from the thermal spike model used in Tyagi's study as well as GO's inhomogeneous nature could also have effects on the defect creation and reduction. These deviations may also be reasons why in this study perpendicular incidence did not leave defects that were detectable in HRTEM. Furthermore, the increase in slit length should follow $1/\tan(\alpha)$, where $\alpha$ is the angle of incidence. However, the average slit lengths derived from the TEM images between 10° and 1° increase by a factor of 3.5, instead of a factor of 10 as it should by the tan relation. This may be attributed to the fact that the TEM grids inside the sample holder had a slightly different angle towards the beam than the calibrated sample holder. Assuming an angle range of 0.5°-3° and 9°-11° (for irradiation under 1° and 10°) gives a factor range of 3-11, which is just in agreement with the measured factor (3.5).

The halo under grazing incidence, compared to the perpendicular geometry has a larger area as seen in Fig. S3. With decreasing angle, the halo becomes even bigger. Consequently, the number of functional groups removed with each impact becomes bigger as well. This is in good agreement with our results that showed that the C/O ratio increases at lower fluences as the angle of irradiation gets smaller. The same uncertainties mentioned for the slit length also come into play for the reduction mechanism.

These simple geometric considerations may explain the shape of the slit pores under grazing incidence and the general trend of increased reduction with decreasing angle of incidence. However, this simplified view on the ion damage is not sufficient to explain the defect mechanism in detail. Further studies with highly accurate angle over fluence measurements in combination with refined simulations are necessary to gain a deeper understanding.

**#2 - Chemical and structural analysis of ion irradiated GO membranes**

Fig. S4A shows SEM images of the surface of a GO membrane, irradiated with 1E12 ions/cm$^2$. The number of ions/cm$^2$ is described as fluence in the following and corresponds to the density of created pores. The surface shows several spots with darker contrast that cannot be found on untreated GO. Enlarging the darker spots reveals, that the spots are created by an agglomeration of slit shaped pores with different lengths, ranging from 10-100 nm.

With X-ray photoelectron spectroscopy (XPS) the carbon to oxygen ratio (C/O ratio) of GO after ion irradiation allows deeper understanding of the chemical changes. Fig 2B shows the increase of the C/O ratio with respect to the fluence with different angles of incidence. For 1 ° the C/O ratio surpasses 3 at 1E10, for 10 ° at 1E11 and for 90° at 1E12. The development of the C1s peak with increasing fluence and variation of the angle of incidence gives further inside into changes of GO induced by ion irradiation. In line with the observed decline in C/O ratio, the C-O fraction of the C1s peak declines with increasing fluence. The fluence necessary to decline the C-O part is less with decreasing angle of incidence. It may also be noted that the ion irradiation seems to mainly reduce the C-O groups while C=O remain more intact as previously observed by the authors of Ref. [34].This reduction of GO by irradiation with ions is also reflected by an increase in conductivity (Fig. S4C). Raman spectroscopy can be used to further analyse the structural and chemical changes after ion irradiation, in particular the $A_D/A_G$ ratio. Fig. S4B shows the results of Raman spectroscopy measurements of the surface of GO in respect to the fluence and angle of incidence. For each sample a map with around 100 measurement points was recorded. The resulting spectra were fitted with the Renishaw WiRE software. Then the area ratios of the respective peaks are averaged over the 100 points. The error bars give the standard deviation. With increasing fluence, the $A_D/A_G$ ratio of GO gradually declines for perpendicular incidence from 2.0 to around 1.5. In contrast to that, the $A_D/A_G$ ratio for grazing incidence declines to 1.24 for a fluence of 1E9. For both 10° and 1° angle of incidence the $A_D/A_G$ ratio rises for fluences

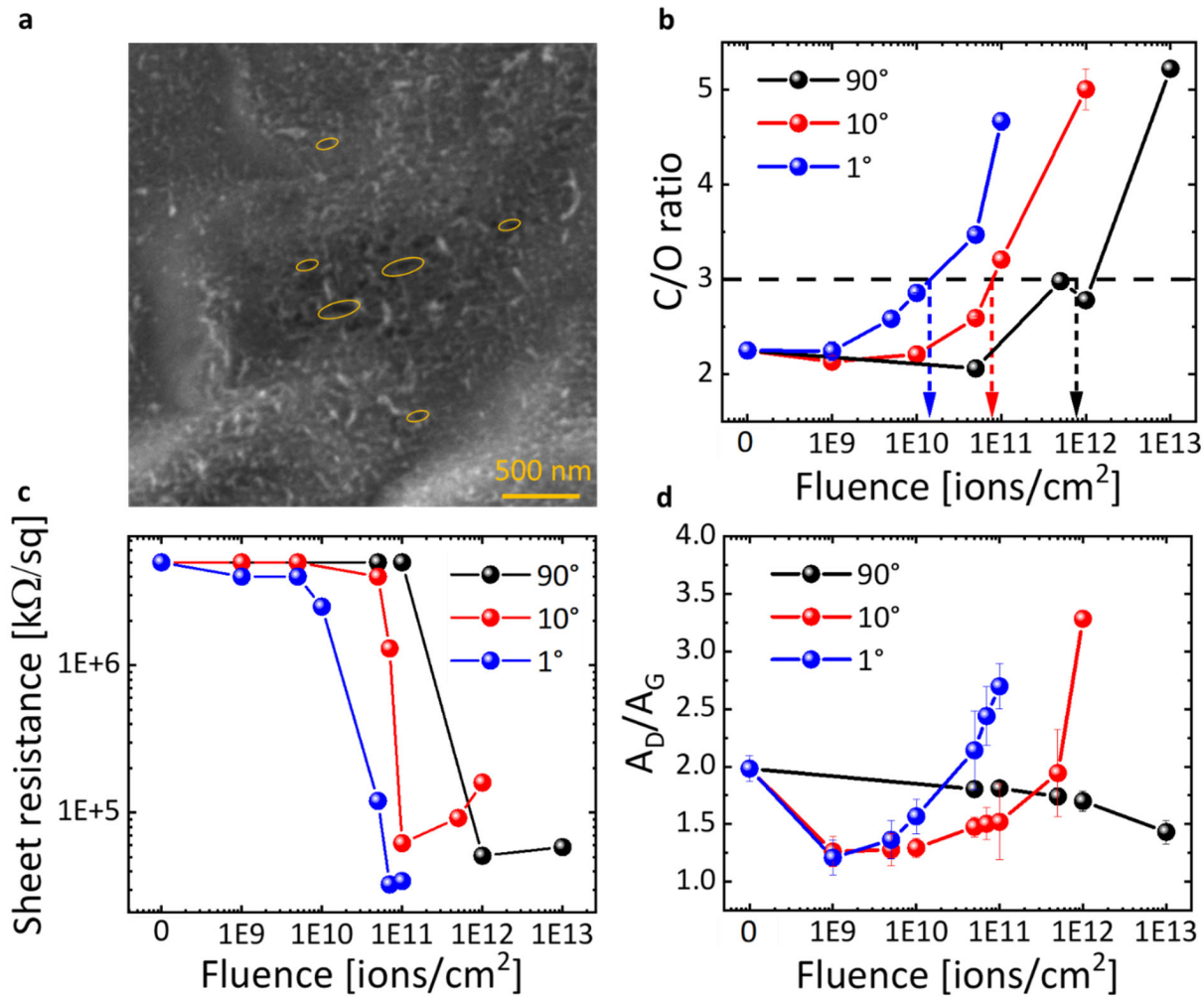

**Fig. S4 Characterization of ion induced pores** (**A**) SEM image of the GO membrane surface, irradiated with 1E12 ions/cm$^2$ under an angle of incidence of 10°. (**B**) C/O ratio of GO membranes in respect to ion fluence and angle of incidence. **c** conductivity of GO membranes in respect to ion fluences and angle of incidence. (**C**) Conductivity changes in respect to fluence and angle of incidence. (**D**) $A_D/A_G$ ratio of GO membranes with different fluences and angle of incidence

higher than 1E9. At 7E10 and 1E12 the $A_D/A_G$ ratio has reached a similar value of unirradiated GO for 1° and 10°, respectively. Then, the value rises for even higher fluences.

For GO this increase in $A_D/A_G$ corresponds to a relative increase in sp$^2$ carbon[45]. However, for perpendicular incidence the $A_D/A_G$ shows a constant decline with respect to fluence despite the observed reduction. This suggests that the perpendicular ion irradiation does induce damage to the sp2 network that was not observable in our TEM study. Furthermore, the development of the interlayer spacing with increasing fluence and angle of incidence was investigated. As shown in fig S6 the interlayer spacing varies within 1 Å after the irradiation but shows no significant trend regarding defect density or angle of incidence.

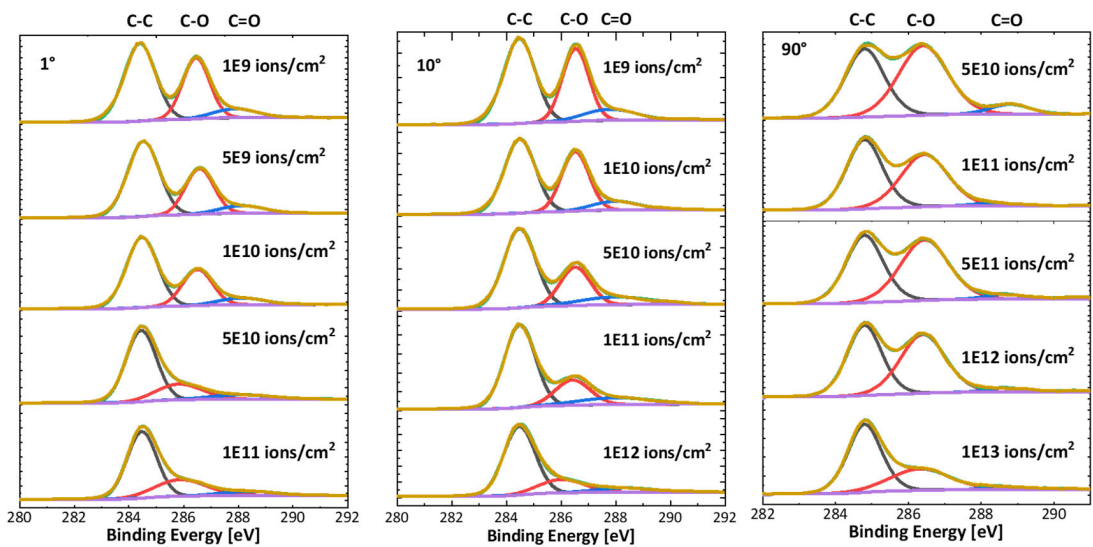

**Fig. S5** C1s spectra of ion irradiated GO under variation of fluence and angle of incidence.

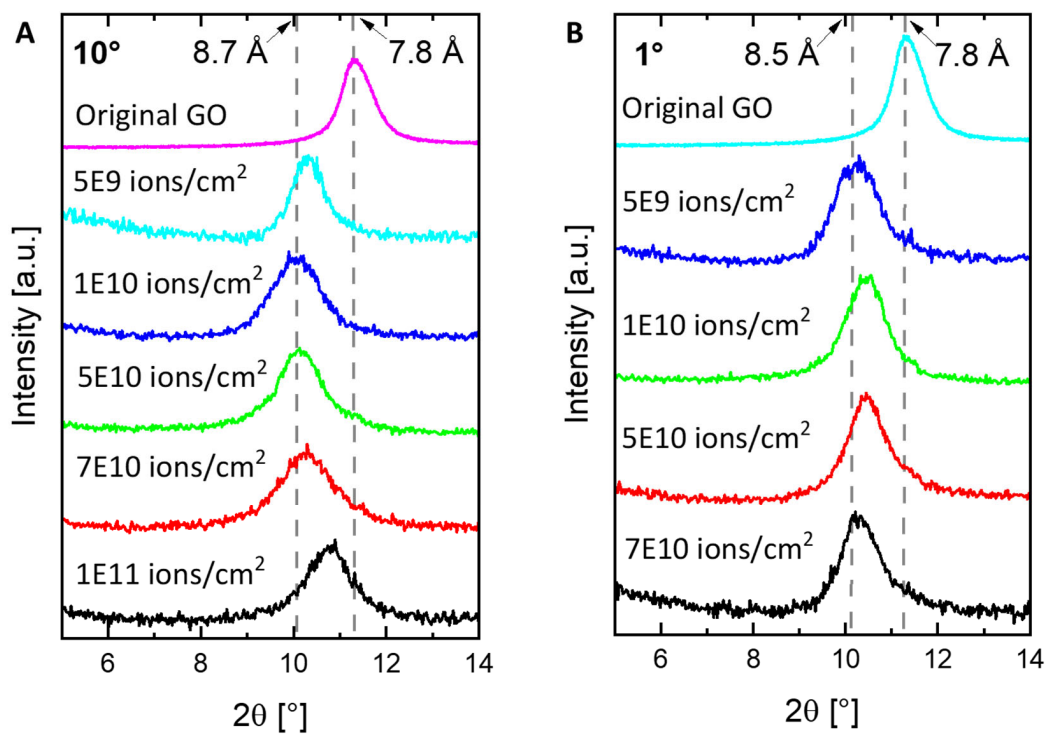

**Fig. S6** XRD pattern of ion irradiated GO membranes.

**#3 - Experimental set-up for mass transport measurements**

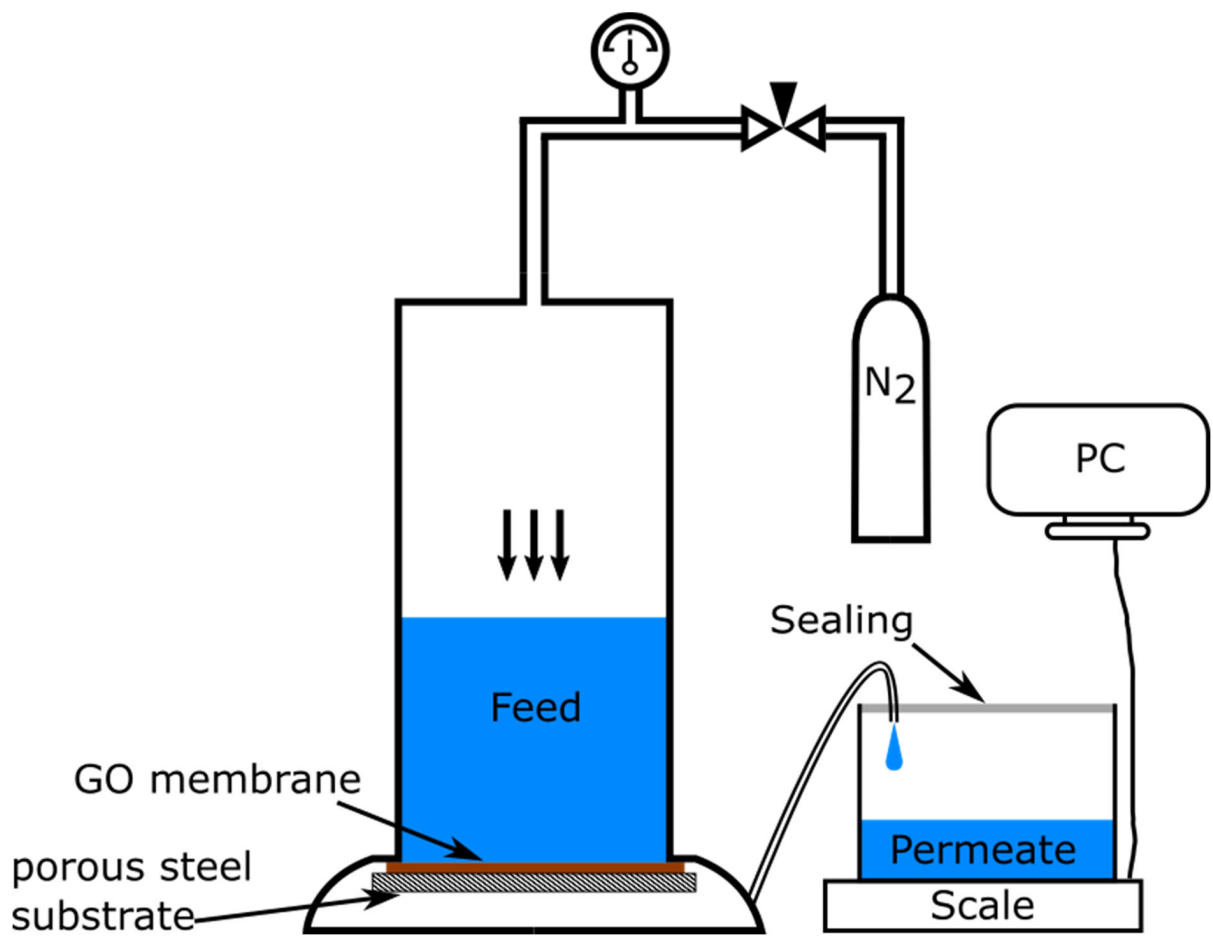

**Fig S7 Experimental set-up for mass transport measurements.** The water and ethanol/water mixture flux were tested in a dead-end cell by Sterlitech. To test the flux, the membranes are mounted into the Sterlitech cell. The membranes are sealed with an O-ring (Viton rubber) to ensure that all mass transport is solely through the membrane. The feed solution is then inserted into the dead-end cell and external pressure is applied via gas pressure. The pressure is measured with a pressure gauge and was held constant at 2 bar for all experiments. The permeate is collected in a sealed vessel placed on a scale. The scale records the water flux. For MB rejection measurements the feed and permeate were analysed via UV-Vis spectrometry to obtain the concentration.

#4 - Thickness of the GO membranes

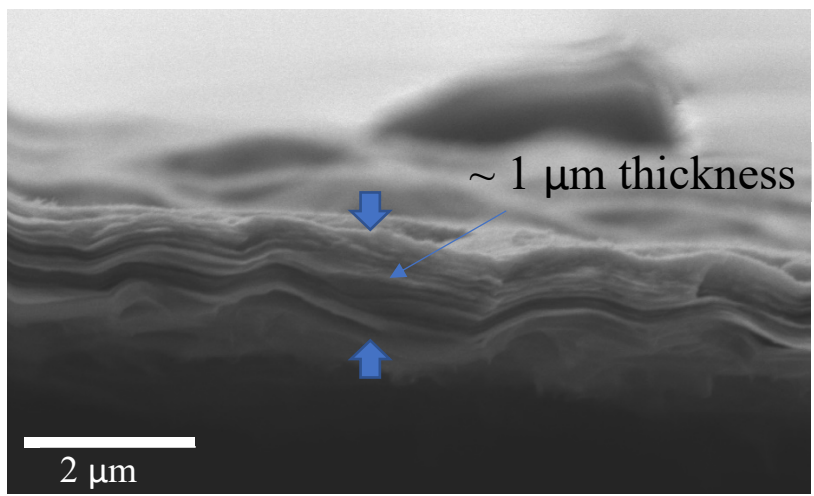

**Fig. S8 SEM Cross section image of graphene oxide membrane recorded with SEM**.

The thickness of the GO membrane is around 1 µm ± 0.1 µm. Throughout the study the membranes were fabricated with the same amount of GO ensuring that the thickness did not play a role. The image shows a GO membrane irradiated with 1E10 ions/cm$^2$. This further confirms that the interlayer structure remains largely intact after ion irradiation.

#5 - Determination of MB rejection rate

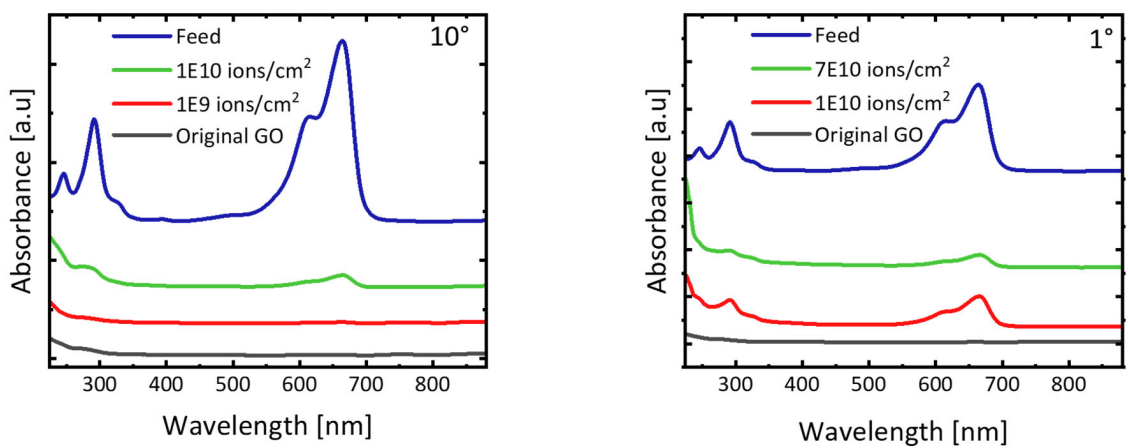

**Fig. S9 UV-Vis pattern of MB rejection test.** Feed and permeate were tested with UV-vis after filtration. Feed side had an initial concentration of 10 ppm. Spectra for feed and permeate filtered with GO membranes irradiated under 10°, 1° with varying fluences.

In order to test the MB concentration of feed and permeate, UV-Vis pattern of the respective solution were recorded. Deconvolutional peak fitting between 600-700 nm allows to determine the peak height of the main absorbance peak of MB at around 667 nm. By recording spectra of standard solutions with concentrations ranging from 0.1-10 ppm, a standard curve is obtained. This allows to find a linear relationship between peak height (absorbance) and concentration according to the Beer-Lambert law. Using the slope of the linear regression then allows to calculate the concentration of an arbitrary MB solution. By this method the concentration of the permeate solution was analysed. From that the MB rejection rates given in Fig. 2 in the main text and Fig S10 were calculated.

#6 - Water flux measurements for GO irradiated under 1°

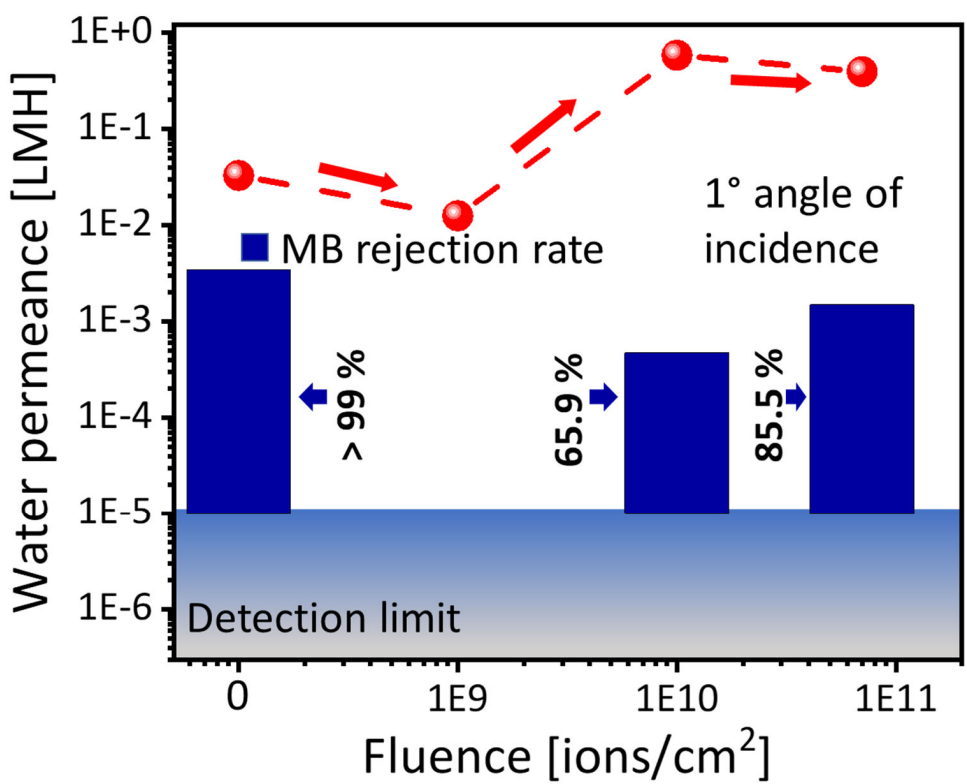

**Fig. S10 Mass transport for ion irradiated GO under 1°.** Red dots show the water flux of GO irradiated under 1° with increasing fluence. The blue histogram shows the MB rejection rate. For 1E9 ions/cm$^2$ the membranes failed before testing for MB rejection.

In addition to the study of GO irradiated under 10°, membranes irradiated under 1° were also investigated. As shown in Fig. S10, the water permeance declines for a fluence of 1E9 ions/cm$^2$ compared to original GO. For higher fluences, 1E10 and 1E11, the water permeance increases up to a factor of ~ 17 compared to original GO. For these high fluences, the MB rejection declines to 65.9 % and 85.5 %. As explained in the main text this means, that a certain proportion of the mass transport only follows the in-plane pores which are big enough to allow the permeance of MB. As irradiation under 1° significantly increases the pore length, it is not surprising that this leads to increased overlap of pores in adjacent layers compared to 10° irradiation. Consequently, the water permeance is increased to a higher level (17x compared to 8x for irradiation under 10°).

**#7 - Additional discussion of MD simulation**

In fig. 3A of the main text, the sudden and rapid permeance of water through the graphitic pores after ~130 ps is marked as a simulation artefact. A closer look on the dynamic of the water molecules around this time frame explains why the permeance was marked as an artefact. Fig. S11A show the histograms of particles during the simulation a certain time steps. The atoms counted as graphene are marked as black, the atoms counted as water molecules are marked as blue. As the dynamic layer of graphene is moved towards the two static layers, pressure on the water reservoir is increased. As shown in fig. S11 the water molecules form two distinct layers under this angstrom-confinement in the water reservoir between the dynamic and first static layer of graphene. The water layers become apparent as two clearly separated peaks in the histogram. When the dynamic layer is moved closer to the static layer the two water layers are pushed closer together. This leads to an increase in pressure and filling of the interlayer space between the two static graphene sheets while permeation through the last static graphene layer is still limited. When the dynamic layer is moved closer to the static layer, the two water layers in the reservoir are forced to combine. As the space between the dynamic layer and static layer is only big enough to host a monolayer of water the excess water is pushed through both pores. Hence, to overcome the barrier for the water to permeate the graphitic pores, a large pressure is necessary that is only reached when there is no more space for the water to form more than one monolayer. Thus,

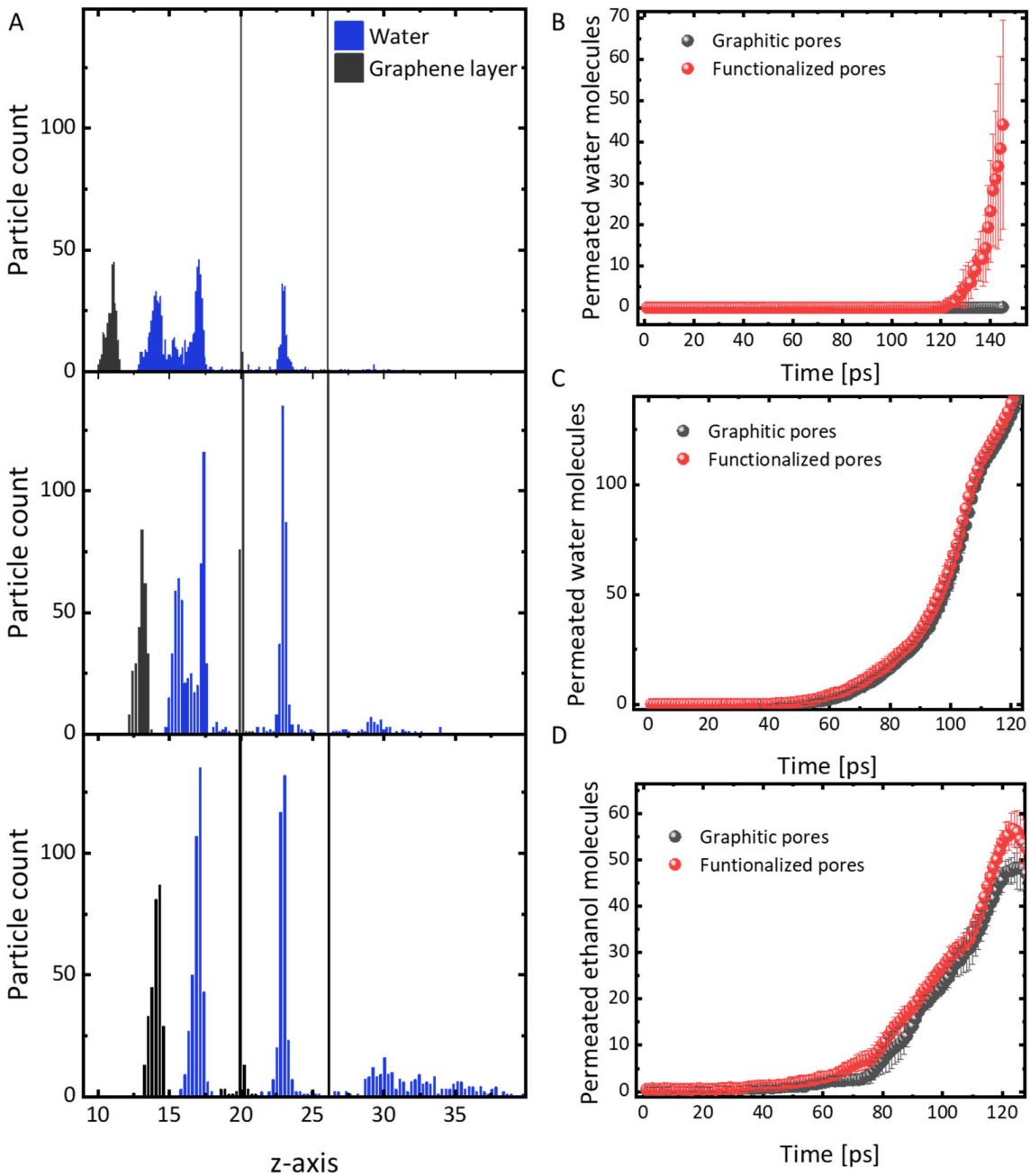

**Fig. S11 Additional MD simulation results and considerations** (A) Dynamic of water molecules (blues bars) in between graphene sheets (black bars). From top to bottom the following time steps are shown (109 ps, 130 ps and 140 ps)

we attribute the rapid and sudden water permeation to this artefact of the simulation rather then free water transport through the graphitic pore.

Fig. S11 B-D show additional MD simulations. Fig. S11B displays the water permeance through a single pore with a smaller diameter of 5.5 Å. This size is typical for intrinsic defects which are common for GO membranes. The simulation shows a clear difference between graphitic and functionalized pores. While functionalized pores allow water transport, graphitic pores show no detectable permeance of water molecules. This further confirms the experimental observation that reduced GO allows less water transport through in-plane pores as this simulation suggests that the mass transport through intrinsic pores is only possible in the presence of functional groups. Additionally, the same simulation was tested for ethanol molecules and no permeation, was detected for both functionalized and graphitic pores. This is in line with the experimental observation that ethanol shows no permeance for original GO.

It may be noted that the pores in the MD simulation are smaller than the ion induced pores in the permeation experiments. Due to computational limitations on the number of atoms and the overall size of the simulation, we had to choose smaller holes in the simulation. With that necessary compromise, we are aiming to compare trends in the simulation and the experiment. In this case the difference in water transport through functionalized and graphitic pores. Even though a direct comparison of the simulations and experiment is not possible like that, the overall trend is the same and should allow us to form the models described in the main text.

Fig. S11C and D display the permeance of water through a single pore with the same pore size as in Fig. 3. However, the permeance in Fig. 3A is measured through two layers of holey graphene. Interestingly, functionalized and graphitic pores show no difference in ethanol and water permeance through a single hole. Consequently, the difference in permeance only becomes apparent with two layers of graphene or more. This underlines the difference between permeation through a single sheet of graphene (one hole) and intralayer transport of angstrom-confined water through in in-plane pores (two holes or more).

Fig. S12 shows the development of water molecules in the space between the static graphene sheets for the permeance presented in fig. 3 of the main text. The configurations at 149 ps are identical to the ones displayed in fig. 3C of the main text. They are replicated for better overview. It becomes apparent, that the water molecules in between the layers with graphitic pores readably form a network of water molecules, which hinders the permeation. At 123 ps only few water molecules permeated the second layer (see fig. 3A) through graphitic pores, whereas functionalized pores already show a steady increase of permeated water molecules. With increasing simulation time, the square water network for graphitic pores becomes denser and almost completely covers the interlayer space except for areas in the proximity of the pores. In contrast, the water molecules in the interlayer space with functionalized pores only form a network towards the end of the simulation period (149 ps). That network, however, only spans over a small part of the interlayer space far away from the pores. Close to the pores water molecules can be found in random orientations, surrounded by an area without water molecules. This indicates that the attraction between the functional groups surrounding the pores and the water molecules disturbs the hydrogen network and promotes water permeance. The time of 123 ps was chosen as an example as the amount of water molecules in the interlayer space of the graphitic pores are same as the amount of water molecules at 149 ps in the interlayer space of functionalized pores. The clear difference in formation of a network in case of graphitic pores compared to the functionalized case with the same number of confined water molecules, underlines that the functional groups disturb the formation of a square water network.

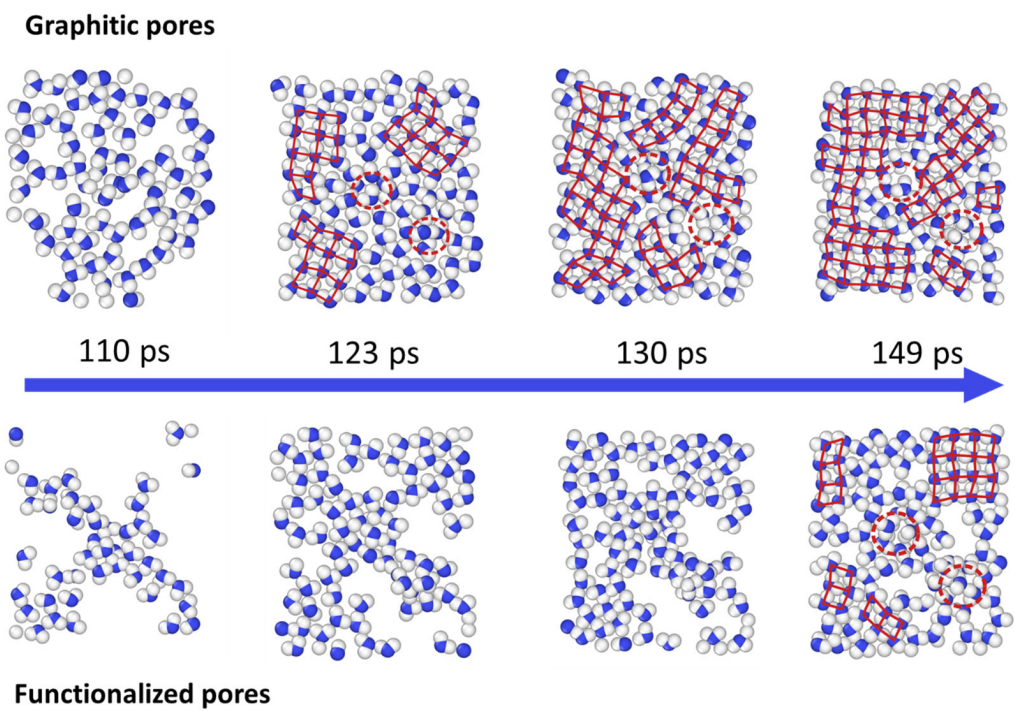

**Fig. S12** Dynamic of water molecules within the interlayer space of static graphene sheets with functionalized and graphitic pores

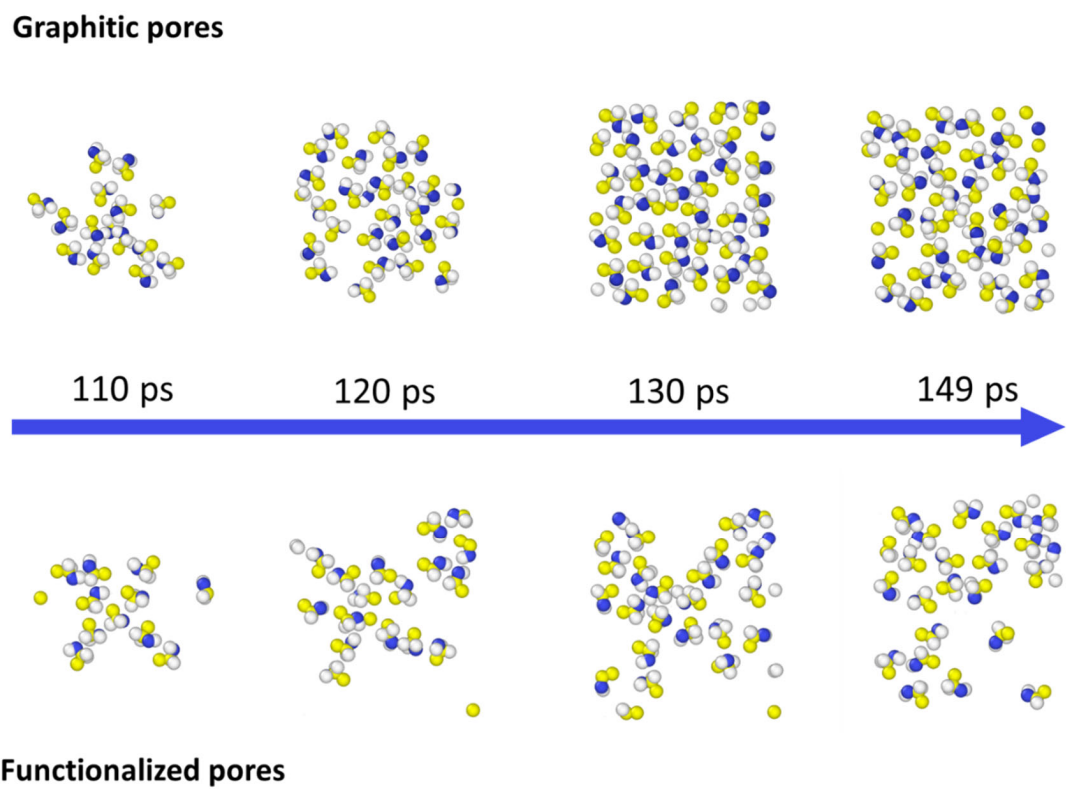

**Fig. S13** Dynamic of ethanol molecules within the interlayer space of static graphene sheets with functionalized and graphitic pores

Figure S13 show the development of ethanol molecules in the interlayer space between the static layers for graphitic and functionalized pores. In both cases no network is apparent. However, the number of molecules stuck in the interlayer space grows to a higher number in the case of graphitic pores compared to functionalized ones. This underlines that the functional groups also facilitate the transport of ethanol through the pores via interaction with the -OH group of ethanol. However, compared to the case of water the lack of a forming network is beneficial for the in-plane pores transport.